\documentclass[letterpaper,11pt]{article}
\usepackage{geometry} %[margin=1in]
\usepackage[utf8]{inputenc}
\usepackage[T1]{fontenc}
\usepackage{lmodern}

\usepackage[dvipsnames,table]{xcolor}
\usepackage{graphicx}
\graphicspath{{./}{figures/}}
\usepackage{hyperref} %[colorlinks]
\hypersetup{colorlinks = true, 
            urlcolor=RoyalBlue,
            citecolor=black,
            }
\usepackage{url}
\usepackage{gensymb}
\usepackage{fancyhdr}
\usepackage{wrapfig}
\usepackage{sidecap}
\usepackage{natbib}  %[square,numbers]
\usepackage{amsmath}
\usepackage{amssymb}
\usepackage{tcolorbox}
\usepackage{longtable}
\usepackage{enumitem}
\usepackage{floatrow}
\newfloatcommand{capbtabbox}{table}[][\FBwidth]
\usepackage{lipsum}
\usepackage[displaymath, mathlines]{lineno}
\modulolinenumbers[2]
\usepackage{float}
\usepackage{pdfpages} %to use \includepdf
\usepackage{todonotes}
\usepackage{sectsty}
\usepackage{authblk}
\usepackage{xspace}

\sectionfont{\fontsize{12}{15}\selectfont}
\subsectionfont{\fontsize{12}{15}\selectfont}

\definecolor{lightgray}{gray}{0.9}
\date{\vspace{-5ex}}

\begin{document}

\newcommand{\apj}{\emph{ApJ}\xspace}
\newcommand{\aap}{\emph{A\&A}\xspace}
% \maketitle

\noindent
\textbf{\large Roman CCS White Paper}

\vspace{2em}
\begin{center}
{\Large Balanced Prism Plus Filter Cadence in \\ the High Latitude Time Domain Survey Core Community Survey}
\end{center}

\vspace{1.5em}
\noindent
\textbf{Roman Core Community Survey:} High Latitude Time Domain Survey

\vspace{0.5em}
\noindent
%Choose from https://hst-docs.stsci.edu/hsp/hubble-space-telescope-call-for-proposals-for-cycle-31/hst-filling-out-the-apt-phase-i-proposal-form
\textbf{Scientific Categories:} stellar physics and stellar types; stellar populations and the interstellar medium; large scale structure of the universe
% TDAMM

\vspace{0.5em}
\noindent
%suggestions for each can be found https://hst-docs.stsci.edu/hsp/hubble-space-telescope-call-for-proposals-for-cycle-31/appendix-b-scientific-keywords
\textbf{Additional scientific keywords:} Supernovae, Exotic Transients, Cosmology, Dark energy

\vspace{1em}
\noindent
\textbf{Submitting Author:}\\
Greg Aldering, Lawrence Berkeley National Lab (galdering@lbl.gov)

\vspace{0.5em}
\noindent
\textbf{List of contributing authors:}\\
% format: Name, affil (email)\\
David Rubin, UH, drubin@hawaii.edu\\
Benjamin Rose, Baylor University (Ben\_Rose@baylor.edu)\\
Rebekah Hounsell, University of Maryland Baltimore County/ NASA Goddard Space Flight Center, (rebekah.a.hounsell@nasa.gov)\\
Saul Perlmutter, University of California, Berkeley (saul@lbl.gov)\\
Susana Deustua, NIST (susana.deustua@nist.gov)\\

\vspace{1em}
\noindent
\textbf{Abstract:} The Nancy Grace Roman Space Telescope's (RST) Wide Field Imager (WFI) is equipped with a slitless prism that can be used for spectroscopic discovery and follow-up of explosive transients at high redshift as part of its High Latitude Time Domain Survey. This is new and unique spectroscopic capability, not only for its original purpose for cosmology, but also for other types of explosive transients. This white paper is intended to help make this new capability more clear to the community. The depth of the RST prism compared to ground-based spectrographs is explored, showing that the RST prism will be unrivaled in the observer-frame NIR. The influence of the selected sky locations on the speed and homogeneity of a RST prism survey is also estimated. This unique new capability should be considered when balancing the HLTDS time devoted to cadenced imaging and spectroscopy.

\thispagestyle{empty}
\newpage
\setcounter{page}{1}

%\url{https://roman.gsfc.nasa.gov/science/ccs_white_papers.html}
%The suggested length is 2 or 3 pages of text, plus figures, tables, and references as needed.  

% \linenumbers
\section{Introduction}

Among the baseline concepts for the "high latitude time domain survey" (HLTDS) as outlined in \cite{Rose2021arXiv211103081R}, there is an advantage to choosing the observing strategy that gives comparable time to prism (P127) and filter observations. Such a balanced prism-plus-filter strategy makes it possible to measure the expansion history of the universe using SNe~Ia, building on the advantages, for both statistical and systematic uncertainties, afforded by spectrum matching of the SNe (\cite{FakhouriTwins2015ApJ...815...58F, 2021ApJ...912...70B, BooneTE2021ApJ...912...71B, Stein_2022ApJ...935....5S}). Such a survey is projected to reach a better statistical-plus-systematics cosmology measurement precision than all-imaging surveys: both options are systematics-limited and the spectroscopy allows better systematics control (and both also achieve comparable statistical-only uncertainty). As \cite{Rubin2022} describes, such a prism+filters survey also makes it possible to measure the redshift of each supernova, avoiding the systematic errors and outliers of photometric redshifts. This balanced prism+filter survey also opens other static spectroscopic science opportunities, using the reconstructed forward-modeled fields that will be observed at $\sim50$ different rotations over the course of the 2-year survey. 

But in addition to these important cosmology measurements made possible by the RST prism, here we want to also emphasize a new potential afforded by a prism survey: the spectroscopic discovery of transients and cadenced spectroscopy for their astrophysical interpretation. Of the many unique aspects of the RST WFI, its wide-field spectroscopic transient search capability is a real breakthrough, as there has never before been such a capability. An interesting advantage of a spectroscopic transient survey is that any fast transients can be characterized immediately from their spectra. With only imaging, one needs to trigger follow-up based on much more limited (potentially only one detection in one filter) information; if such a trigger requires space resources, it could well miss the subsequent evolution of such events while still bright\footnote{For example, in JWST Cycle~2, only 8 target-of-opportunity triggers with lead-time of less than 14 days were allowed.}. With its $R \sim 100$ spectral resolution, corresponding to a restframe photospheric velocity of 3000 km/s, the prism observations will have significantly higher sensitivity for fast explosive transients than the grism that is being employed for the High Latitude Wide Area Survey, and will be cadenced. This potential is important to consider when designing the HLTDS. We suggest that the prism+imaging survey variant offers both the most systematics control for the SN cosmology measurements and the most flexibility and balance for transient science. 

% JWST allowed 8 disruptive ToOs in Cycle 2
%Non-disruptive ToO: A ToO for which the activation time is greater than 14 days
%Disruptive ToO: A ToO for which the activation time is less than 14 days and greater than 3 days
%Ultra-disruptive ToO: A ToO for which the activation time is less than than 3 days

\vspace{1em}\noindent
\section{Comparison to Conventional Ground-based Spectroscopic Follow-Up}

The typical mode of operation with transients is to discover them with imaging and determine their astrophysical properties from follow-up spectroscopy. Here we show that RST prism observations surpass ground-based spectroscopy in the observer-frame NIR for spectroscopic follow-up.

\begin{figure}
    \centering
    \includegraphics[width=0.8\textwidth]{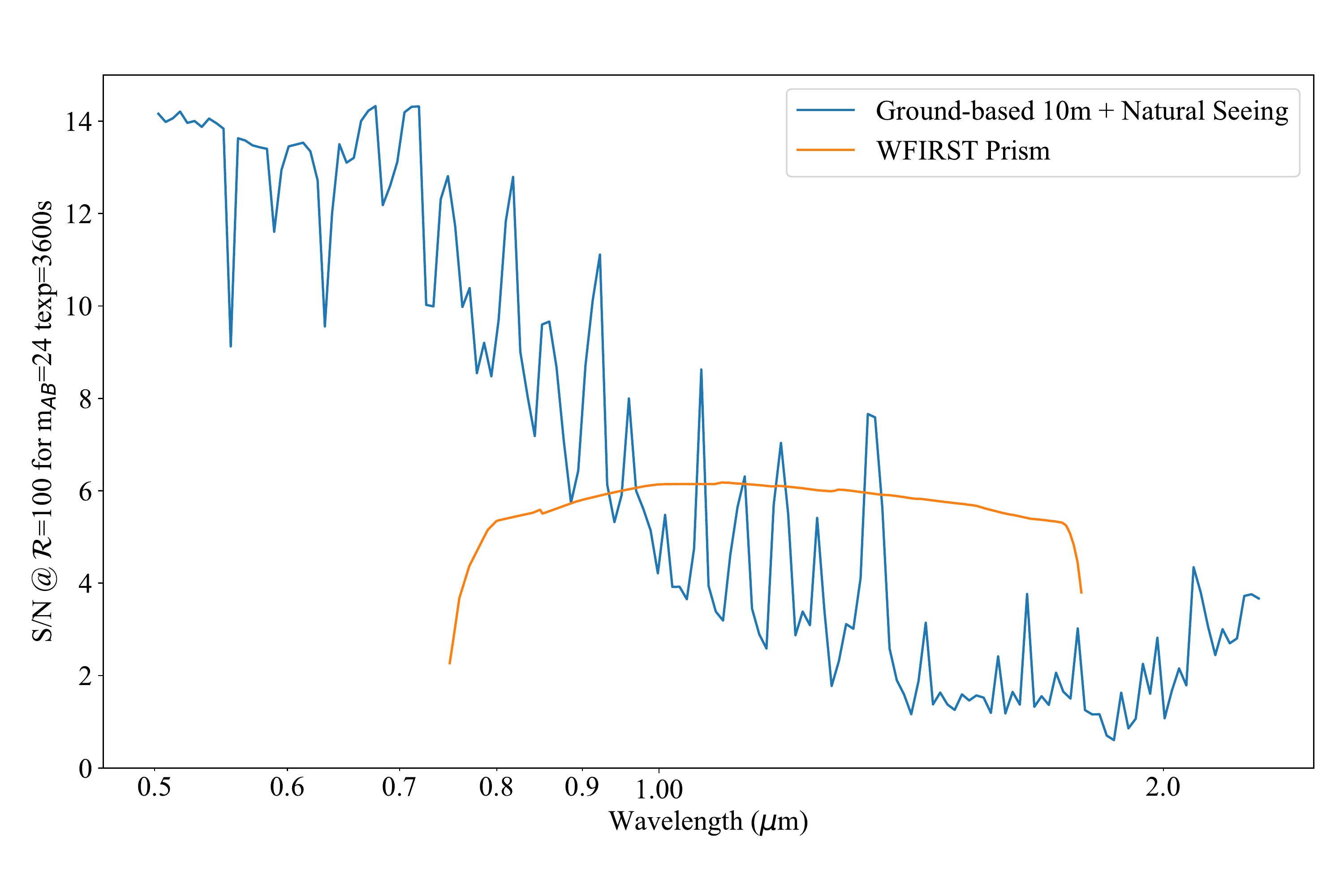}
    \caption{S/N comparison of RST prism with hypothetical ground-based spectrograph covering observer-frame wavelengths of 0.4--2.3\,$\mu$m with a throughput of 25\% for a flat-spectrum $m_{AB}=24$ source observed for 1\,hr. The data are binned to a resolution of 100, equivalent to a velocity resolution of 3000\,km/s, as would be done to find features in extragalactic explosive transients. In the range 1.0--1.8\,$\mu$m, the RST prism is clearly superior. With the log-wavelength scale used here, a redshift corresponds to a shift in the $x$ direction. It also displays features at fixed velocity resolution. Thus, it is most appropriate for studying the sensitivity for explosive transients at unknown redshifts.}
    \label{fig:snr}
\end{figure}

In Figure\,\ref{fig:snr} the S/N for the RST prism is compared to observations with a hypothetical instrument on a ground-based 10-m telescope having an end-to-end throughput of 25\%, and clear, dark skies and seeing of 0.6\,arcsec. The hypothetical target has a flat-spectrum with brightness $m_{AB}=24$, and is observed for 1\,hr. Even compared to such an idealized instrument, the RST prism performs significantly better in the wavelength range 1.0--1.8\,$\mu$m.\footnote{Recall that for background-limited observations, the exposure time needed to catch up to the RST prism would go as the square of the ratio of S/N. So to get from S/N$\sim 1.5$ up to S/N$\sim 5.5$, as at 1.5\,$\mu$m in Figure\,\ref{fig:snr}, requires 13$\times$ as much exposure time.} Moreover, the RST prism sensitivity is very uniform, whereas that from the ground is very non-uniform due to H$_2$O absorption and OH emission. The non-uniform sensitivity with wavelength translates to non-uniformity with respect to redshift and with respect to key spectra features. Compared to this idealized ground-based spectrograph, existing spectrographs with comparable simultaneous wavelength coverage (e.g., X-shooter) have much lower throughput, while spectrographs with comparable throughput typically have non-simultaneous coverage of, e.g., $Y$, $J$, $H$, $K$ (e.g., MOSFIRE) in the near-infrared. Gains from using adaptive optics require Strehl ratios much better than delivered today.\footnote{See ``The Case for an AO-enabled Faint Object Integral Field Spectrograph'' at \\
\url{https://www2.keck.hawaii.edu/inst/scistratplan/whitepapers.html}}
%%https://
%www2.keck.hawaii.edu/inst/scistratplan/whitepapers/The Case for an AO\_enabled Faint Object Integral Field Spectrograph.pdf}.

In addition, there will be a significant fraction of the time when optimal follow-up of a transient detected in an RST CSV field is not possible --- due to daylight, bright Moon, unfavorable airmass, instrument unavailable, bad, weather or poor seeing --- from the ground. Even then, it often will be challenging to detect the target with the spectrograph acquisition system, at least with sufficient astrometric accuracy to guarantee it will be centered in the spectrograph slit before commencing a long exposure. 

Further in favor of the RST prism, the observations desired for SN~Ia cosmology will be cadenced and have depth similar to that simulated here \citep{Rubin2022}. Such a combination of depth and cadence observations will --- for the first time --- enable spectroscopic {\it discovery} of high-redshift transients. In contrast to occasional, and inhomogeneous ground-based spectroscopy, the RST prism will be able to obtained cadenced follow-up, enabling cross-epoch co-addition for greater depth, or monitoring of spectral evolution with time.

Other surveys, like LSST, will produce transient discoveries for brighter transients, and ground-based telescopes will be more capable of obtaining spectroscopy for them. It is for faint transients in the observer-frame NIR where the RST prism survey will truly shine.

\vspace{1em}\noindent
\section{Survey Speed and Homogeneity}

In order to maximize the prism survey effectiveness it should be conducted in regions with low zodiacal background. This is because each pixel on the WFI detector receives the full spectral range of zodiacal light, and this is enough to make practical P127 exposures background-limited (rather than readout-noise limited). For faint targets, this means that the survey speed is inversely proportional to the zodiacal light background for exposure times more than a few minutes. In addition, away from the ecliptic poles the zodiacal light level will have a predictable annual variation. In order to ensure survey homogeneity, this annual background variation should be minimized by staying close to the ecliptic poles, or the exposure times should be adjusted to compensate.

In another Roman white paper, the locations of potential HLTDS fields are discussed, with an emphasis on existing deep fields, and having low Galactic extinction\footnote{A driver for SN Cosmology}. Here we use the \texttt{ZodiPy} \citep{ZodiPy20222} to compare the survey speed and homogeneity of these fields within the RST continuous viewing zone based on their zodiacal light level compared to the darkest good field (the IRAC Dark Field) and the fractional range in that level over the course of a year. Table\,1 shows the results. After the first five fields the survey speed for a prism survey drops quickly and the inhomogeneity increases substantial (i.e., the first 5 fields have inhomogeneity around 20\% across the year, but then it rises to 40\% or much more for fields with lower speeds). Ecliptic latitude ($\epsilon$) also serves as a pretty good guide, but again, above $|\epsilon| > 75^\circ$, Galactic extinction and stellar interference become worse.

Reduced survey speed and increasing inhomogeneity can be compensated with longer exposure times, but only up to a point. There will be systematic sky subtraction errors at some level, and these will get worse relative to target brightnesses as the level of the zodiacal light increases. Thus, selecting fields with low zodiacal background (but also low Galactic extinction, $A_V$, as indicated in Table\,1) is critical for the prism survey.

\begin{table}
\begin{center}
{Table 1: Prism Survey Relative Speed and Inhomogeneity}
\vskip 5pt
\begin{tabular}{lrrrccccc}
\hline \\[-0.8em]
Name  & \multicolumn{1}{c}{$\epsilon$} & \multicolumn{1}{c}{RA} & \multicolumn{1}{c}{Dec} &  Relative  & Relative & $A_V$ &   In \\
      &       &       &       &  Speed &   Inhomo-   &       & CVS \\
      & (deg) & (deg) & (deg) &        &   geneity   & (mag) &      \\[0.3em]
\hline \\[-0.7em]
IRAC Dark Field & $ +86.92$ &  265.00 & $  +69.0$ &   1.000 &   0.190 &    0.12 & True \\
SNAP-N          & $ +75.41$ &  246.25 & $  +57.0$ &   0.957 &   0.223 &    0.02 & True \\
ELAIS N-1       & $ +72.63$ &  242.75 & $  +55.0$ &   0.939 &   0.252 &    0.02 & True \\
ADFS            & $ -72.96$ &   71.00 & $  -52.3$ &   0.930 &   0.202 &    0.02 & True \\
SNAP-S          & $ -71.67$ &   67.50 & $  -52.0$ &   0.921 &   0.199 &    0.02 & True \\
ELAIS N-2       & $ +62.52$ &  251.70 & $  +41.0$ &   0.848 &   0.608 &    0.04 & True \\
LSST 820        & $ -61.83$ &  119.56 & $  -43.4$ &   0.827 &   0.770 &    0.94 & True \\
EGS             & $ +59.79$ &  214.25 & $  +52.5$ &   0.820 &   0.397 &    0.02 & True \\
Goods-N         & $ +57.29$ &  189.19 & $  +62.2$ &   0.791 &   0.415 &    0.03 & True \\
Deep2A          & $ +56.83$ &  253.00 & $  +34.9$ &   0.783 &   0.837 &    0.05 & True \\
Bootes          & $ +45.98$ &  218.00 & $  +34.3$ &   0.643 &   1.179 &    0.04 & False\\
SPT Deep        & $ -46.20$ &  352.50 & $  -55.0$ &   0.639 &   1.170 &    0.03 & False\\
Lockman Hole    & $ +45.26$ &  161.25 & $  +58.0$ &   0.633 &   1.045 &    0.03 & False\\
CDFS            & $ -45.19$ &   53.12 & $  -27.8$ &   0.624 &   1.211 &    0.02 & False\\
ELAIS S-1       & $ -42.63$ &    8.75 & $  -43.7$ &   0.590 &   1.386 &    0.02 & False\\
LSST 858        & $ -35.51$ &  187.62 & $  -42.5$ &   0.477 &   3.202 &    0.25 & False\\
LSST 1200       & $ -31.42$ &  176.63 & $  -33.1$ &   0.416 &   4.229 &    0.21 & False\\
XMM-LSS         & $ -18.37$ &   37.83 & $   -4.5$ &   0.233 &  11.296 &    0.07 & False\\
VVDS14h         & $ +16.15$ &  210.00 & $   +5.0$ &   0.204 &  15.137 &    0.07 & False\\
SSA22           & $ +10.32$ &  334.25 & $   +0.4$ &   0.117 &  47.766 &    0.15 & False\\
LSST 1689       & $  +9.38$ &  201.85 & $   +0.9$ &   0.108 &  50.332 &    0.07 & False\\
COSMOS          & $  -9.36$ &  150.12 & $   +2.2$ &   0.101 &  57.847 &    0.05 & False\\
Deep2B          & $  +2.98$ &  352.50 & $   +0.0$ &   0.025 & 797.237 &    0.12 & False\\
%Deep2B          & $  +2.98$ &  352.50 & $   +0.0$ &   0.025 & 797.237 &    0.08 & False\\
\hline
\end{tabular}
\end{center}
\end{table}

\section{Conclusion}

Here we have provided a general outline of capabilities of an RST prism survey. We have purposely avoided making projections about which types and numbers of known transients will be discovered. The power of the RST prism survey hopefully will lead to some surprises in this arena. The RST prism's forte will be faint NIR transients, beyond the reach of, e.g., LSST and ground-based spectrographs. Exploiting this capability would require that a substantial portion of the HLTDS observing time - comparable to the imaging portion - be dedicated to RST prism observations (as outlined in \citet{Rose2021, Rubin2022}). The choice of field location proves to be critical to how much time is actually needed, with fields at high ecliptic latitude but low Galactic extinction offering the best combination of survey power and speed. 

%Because this unique new capability might not be readily apparent, we want to bring it to the attention of the RST project when balancing the HLTDS time devoted to cadenced imaging and spectroscopy.

\bibliographystyle{plainnat}
\bibliography{library}

\end{document}